\documentclass[nosuperscriptaddress,amsmath,amssymb,prd,nofootinbib,twocolumn]{revtex4-2}

\usepackage{mathrsfs}
\usepackage[x11names]{xcolor}
\usepackage{graphicx}
\usepackage{dcolumn}
\usepackage{bm}
\usepackage[pdftex]{hyperref}
\usepackage{soul}
\usepackage{float}
\usepackage{orcidlink}
\usepackage{enumitem}
\usepackage[skip=5pt, indent=10pt]{parskip}

\hypersetup{pdftitle={Deciphering the Sources of Cosmic Neutrinos},
pdfsubject={Deciphering the Sources of Cosmic Neutrinos},
pdfauthor={Groth \& Ahlers},
pdfstartview={FitH},
colorlinks=true,
bookmarksopen=false,
bookmarksnumbered=false,
bookmarksopenlevel=0,
linkcolor=Blue1!60!black,
citecolor=Green1!50!black,
urlcolor=Blue1!70!black
}

\begin{document}

\title{Deciphering the Sources of Cosmic Neutrinos} 

\author{Kathrine Mørch Groth\,\orcidlink{0000-0002-1581-9049}}
 \email{kathrine.groth@nbi.ku.dk}
\affiliation{Niels Bohr Institute, University of Copenhagen, Blegdamsvej 17, 2100 Copenhagen, Denmark}

\author{Markus Ahlers\,\orcidlink{0000-0003-0709-5631}}
\affiliation{Niels Bohr Institute, University of Copenhagen, Blegdamsvej 17, 2100 Copenhagen, Denmark}

\begin{abstract}
More than a decade ago, the IceCube Neutrino Observatory discovered a diffuse flux of 10~TeV -- 10~PeV neutrinos from our Universe. This flux of unknown origin most likely emanates from an extragalactic population of neutrino sources, which are individually too faint to appear as bright emitters. We review constraints on extragalactic neutrino source populations based on the non-detection of the brightest neutrino source. Extending previous work, we discuss limitations of source populations based on general neutrino luminosity functions. Our method provides more conservative but also statistically more robust predictions for the expected number of observable sources. We also show that the combined search of the brightest neutrino sources via weighted stacking searches or the analysis of non-Poissonian fluctuations in event-count histograms can improve the discovery potential by a factor of 2-3 relative to the brightest source.
\end{abstract}

\maketitle

\section{\label{sec:1}Introduction}

A major milestone in neutrino astronomy was achieved in 2013, when IceCube reported the first observation of a diffuse flux of astrophysical neutrinos in the $10$~TeV--$10$~PeV energy range~\cite{Aartsen:2013bka,Aartsen:2013jdh,Aartsen:2014gkd}. Since then, the signal has been studied in multiple complementary analyses characterizing the spectrum and neutrino flavour composition~\cite{IceCube:2018pgc,IceCube:2020wum,IceCube:2020acn,IceCube:2021uhz,Abbasi:2021qfz,IceCube:2024fxo,IceCube:2020fpi,IceCube:2024nhk}. Neutrino emission at the observed flux level has been predicted from a variety of source classes; see \textit{e.g.} Refs.~\cite{Ahlers:2018fkn,Vitagliano:2019yzm,Kurahashi:2022utm} for reviews. Various follow-up studies have tried to identify the sources responsible for this diffuse neutrino emission, including $\gamma$-ray bursts~\cite{IceCube:2017amx,Abbasi:2022whi,IceCube:2023woj}, $\gamma$-ray blazars~\cite{Aartsen:2016lir,Abbasi:2022uox}, active galaxies~\cite{IceCube:2021pgw,IceCube:2023htm,IceCube:2024ayt,IceCube:2024dou,Abbasi:2024ewg}, galaxy clusters~\cite{Abbasi:2022uqr}, starburst galaxies~\cite{IceCube:2021waz}, supernovae~\cite{IceCube:2023esf}, Galactic novae~\cite{IceCube:2022lnv}, pulsar wind nebulae~\cite{IceCube:2020svz} or X-ray binaries~\cite{IceCube:2022jpz}. All of these analyses are thus far inconclusive.

In parallel, IceCube has made some progress in the identification of candidate neutrino sources in recent years. These include the observation of neutrinos in coincidence with the $\gamma$-ray blazar TXS 0506+056 \cite{IceCube:2018cha,IceCube:2018dnn} and the Seyfert II galaxy NGC 1068~\cite{IceCube:2019cia,IceCube:2022der}. High-energy neutrino alerts coincident with tidal disruption events have also been observed~\cite{Stein:2020xhk,Reusch:2021ztx}. However, none of these analyses have yielded conclusive discoveries of neutrino point sources at the 5$\sigma$-significance level~\cite{IceCube:2015usw,IceCube:2016tpw,IceCube:2018ndw,IceCube:2019cia,IceCube:2020nig,IceCube:2025zyb}. More recently, IceCube has also observed a neutrino signal from our own Milky Way Galaxy consistent with the expected glow from cosmic-ray interactions in the interstellar medium~\cite{IceCube:2023ame,Ambrosone:2023hsz}. Despite this progress, the origin of the extragalactic diffuse neutrino flux remains uncertain.

The upper limit on neutrino emission from individual point sources has implications for the properties of candidate neutrino sources responsible for the observed diffuse flux~\cite{Lipari:2008zf,Silvestri:2009xb,Murase:2012df,Ahlers:2014ioa,Kowalski:2014zda,Mertsch:2016hcd,Murase:2016gly,Neronov:2018wuo,Ahlers:2018fkn,IceCube:2018omy,Yuan:2019ucv,Palladino:2019hsk,IceCube:2018ndw,IceCube:2019xiu,Ackermann:2019ows,Capel:2020txc,IceCube:2022ham,Fiorillo:2023mjr, Das:2020hev}. A common approach is to approximate the source population as a collection of standard candles whose comoving density evolves with redshift. With this approach, candidate neutrino source populations can be parametrized in terms of their typical luminosity and local density. Source populations of rare but bright neutrino sources are more likely to be visible as bright individual sources compared to those that are numerous but faint.

In this paper, we explore strategies to constrain extragalactic neutrino source populations by leveraging both the observation of the high-energy diffuse flux and the absence of detections of individual point sources. We start from the dependence of the combined diffuse neutrino emission and the flux of individual point sources on a general neutrino luminosity function. In the special case of neutrino standard candles, this allows us to discuss exclusion limits in terms of neutrino luminosity $L_\nu$ and local density $\rho_0$. In contrast to previous studies, we base our discussion on the flux distribution of sources, rather than targeting the expected flux of the brightest source. We argue that this leads to statistically more robust exclusion contours of the $L_\nu$-$\rho_0$ parameter space and more conservative limits.

We discuss candidate populations of steady neutrino sources in terms of their neutrino luminosity function motivated by multi-messenger relations. Extending previous work, we extract the expected number of observable sources from their luminosity functions and its implications for the contribution of these sources to the diffuse flux. We also show that a naive application of the standard-candle approximation can lead to misleading conclusions in some cases, such as BL Lacertae objects (BL Lacs).

We illustrate how existing neutrino data can already improve constraints on standard-candle populations based on only the brightest source. We derive the expected flux distribution of nearby extragalactic sources distributing as in Euclidean space and show that source catalogue analyses using weighted stacking or non-Poissonian fluctuations in event-count histograms can improve luminosity limits of standard-candle populations by an order of magnitude.

The paper is organized as follows. We start in Section~\ref{sec1} by summarizing the relation of the observed diffuse neutrino emission to the luminosity function characterizing the steady neutrino emission of extragalactic populations. In Section~\ref{sec2} we discuss the expected point-source discoveries for general source populations and its implications for the case of standard-candle populations. In Section~\ref{sec3} we illustrate our method for various candidate populations of multi-messenger sources. We then discuss improvements that can be expected by source catalogue analyses in Section~\ref{sec4} before concluding in Section~\ref{sec5}.

\section{\label{sec1} Diffuse Neutrino Emission}

As our working hypothesis, we assume in the following that the diffuse neutrino flux is the result of steady neutrino emission from extragalactic sources. An individual source of the population isotropically emits neutrinos at a spectral rate $Q_\nu(E_\nu) \propto E_\nu^{-\gamma}$ (in units of ${\rm GeV}^{-1}{\rm s}^{-1}$). We will assume that the spectral index $\gamma$ of the neutrino emission is universal, so that the neutrino emission from individual sources can be parametrized in terms of its monochromatic neutrino luminosity $L_{\nu}$ (in units of ${\rm erg}/{\rm s}$ and summed over flavour) at a pivot energy of $100$~{\rm TeV}:
\begin{equation}
L_\nu \equiv [E_\nu^2Q_\nu(E_\nu)]_{E_\nu = 100\,{\rm TeV}}\,.
\end{equation}
The neutrino source population is then described by a neutrino luminosity function (LF) ${d^2N}/{dL_{\nu} dV_c}$ which gives the number of sources per comoving volume $V_c(z)$ within the luminosity interval $[L_\nu,L_\nu + dL_\nu]$. 

The combined angular-averaged diffuse neutrino flux $\Phi_\nu$ (in units of ${\rm GeV}^{-1}{\rm s}^{-1}{\rm cm}^{-2}{\rm sr}^{-1}$) from a population of sources is then given by (see, {\it e.g.}~Ref.~\cite{Ahlers:2018fkn}):
\begin{equation}\label{eq:E2Phi}
[E_\nu^2 \Phi_{\nu}]_{100\,{\rm TeV}} = \frac{c}{4\pi}  \int dz  \frac{(1+z)^{-\gamma}}{H(z)} \mathcal{E}(z),
\end{equation}
where $H(z)$ is the Hubble parameter at redshift $z$ and $\mathcal{E}(z)$ is the luminosity density or {\it emissivity} (in units of ${\rm erg}\,{\rm s}^{-1}\,{\rm Gpc}^{-3}$) defined as:
\begin{equation}\label{eq:emissivity}
\mathcal{E}(z)  \equiv
 \int dL_{\nu} L_{\nu} \frac{d^2N}{dL_{\nu} dV_c}(L_{\nu}, z)\,.
\end{equation}   
We will assume the concordance model of cosmology with local Hubble radius $R_H \equiv c/H_0 \simeq4.45\,{\rm Gpc}$, dark energy density $\Omega_{\Lambda}=0.685$, and matter density $\Omega_{m}=1-\Omega_{\Lambda}$~\cite{ParticleDataGroup:2024cfk}.
It is convenient to account for the redshift dependence of the diffuse flux by the dimensionless quantity:
\begin{equation}\label{eq:xi_F}
 \xi_z \equiv \int_0^{\infty}   dz  \frac{(1+z)^{-\gamma}}{\sqrt{\Omega_{\Lambda}+ (1+z)^3\Omega_m}} \, \frac{\mathcal{E}(z)}{\mathcal{E}(0)}\,,
\end{equation}
which accounts for the redshift evolution of the local emissivity $\mathcal{E}_0\equiv\mathcal{E}(0)$. The diffuse flux can then be written in the compact form:
\begin{equation}
[E_\nu^2 \Phi_{\nu}]_{100\,{\rm TeV}} = \frac{R_{H}}{4\pi}  \xi_z\mathcal{E}_0\,,
\end{equation}
where $\xi_z$ corrects the {\it naive} expectation of a static Universe with finite size $R_H$.

While the diffuse astrophysical flux has been observed by IceCube with high significance, its spectral shape remains uncertain. Various complementary analyses agree on the flux level at a pivot neutrino energy of $100\,{\rm TeV}$ within a factor of 2, while the best-fit spectral index falls into the range of 2.3--2.9~\cite{IceCube:2018pgc,IceCube:2020wum,IceCube:2020acn,IceCube:2021uhz,Abbasi:2021qfz,IceCube:2024fxo}. For concreteness, we will normalize the diffuse flux (per-flavour and summed over neutrinos and anti-neutrinos) to the result of the recent study~\cite{IceCube:2024fxo}:
\begin{equation}\label{eq:E2PhiIC}
[E^2  \Phi^{\rm IC}_{\nu}]_{100\,{\rm TeV}} = \left(1.68^{+0.19}_{-0.22}\right) \times 10^{-8} \frac{\rm GeV}{ {\rm cm}^{2}\,{\rm s}\,{\rm sr}}\,,
\end{equation}
with a best-fit spectral index of $\gamma = 2.58^{+0.10}_{-0.09}$. The value (\ref{eq:E2PhiIC}) fixes the effective redshift integrated emissivity $\xi_z\mathcal{E}_0$ that is required to account for 100\% of the observed total diffuse flux (summed over flavour) to:
\begin{equation}\label{eq:emissivityIC}
\xi_z\mathcal{E}^{\rm IC}_0 = \left(2.2^{+0.2}_{-0.3}\right)\times10^{45}\,{\rm erg}\,{\rm s}^{-1}\,{\rm Gpc}^{-3}\,.
\end{equation}
This value (including its 68\% C.L.) is indicated as the bold horizontal line in the two plots of Fig.~\ref{fig:rhoL}.

So far, our discussion assumed general neutrino LFs with a uniform spectral index $\gamma$. We now consider neutrino LFs, which are separable as:
\begin{equation}\label{eq:separable}
\frac{d^2N}{dL_{\nu} dV_c}(L_{\nu}, z) = \rho(z)f(L_{\nu})\,,
\end{equation}
where $\rho(z)$ is the density of sources per comoving volume and $f(L_{\nu})$ the probability distribution of finding a neutrino source with luminosity $L_{\nu}$. In this case, the local emissivity becomes a product of local density and average luminosity, $\mathcal{E}_0 = \rho_0 \langle L_\nu\rangle$, and the redshift evolution factor in Eq.~(\ref{eq:xi_F}) is identical to the one used in earlier studies, {\it e.g.}~Ref.~\cite{Ahlers:2018fkn}.

For concreteness, we assume in the following that the comoving source density follows the simple form:
\begin{align}
\rho(z) &= \rho_0 \, 
\begin{cases}
(1+z)^m, \quad z < z_{\rm br}, \\
(1+z_{\rm br})^m, \quad  z_{\rm br} < z < z_{\max},
\end{cases}
\label{eq:rho}
\end{align}
with power index $m$ up to a break redshift $z_{\rm br} = 1.5$ and a maximum redshift $z_{\rm max}=6$. We consider three benchmark cases: {\it a)} no evolution with $m=0$, {\it b)} evolution following the star formation rate (SFR) with $m=3$, and {\it c)} strong evolution with $m=5$. 
The corresponding redshift evolution factors~(\ref{eq:xi_F}) for a spectral index of $\gamma=2$ ($\gamma=3)$, are $\xi_z\simeq0.5$ (0.4) for the case of no evolution, $\xi_z\simeq2.7$ (1.3) for SFR evolution and $\xi_z\simeq11.7$ (4.8) for strong evolution.

\section{\label{sec2} Point-Source Limits}

We now turn to the identification of individual members of an extragalactic population of neutrino sources. Due to the limited angular resolution of neutrino telescopes, these sources are expected to be unresolvable point sources (PSs) that can be observed as event clusters within the angular resolution on top of a large atmospheric background.
The expected spectral flux per PS (in units of ${\rm GeV}^{-1}\,{\rm cm}^{-2}\,{\rm s}^{-1}$ and summed over flavour) is given as:
\begin{equation}\label{eq:PS}
\phi_{\nu}^{\mathrm{PS}} (E_\nu) = \frac{(1+z)^2}{4\pi d_L^2(z)} \,  Q_{\nu} ((1+z)\, E_\nu)\,,
\end{equation}
where $d_L(z)$ is the luminosity distance at redshift $z$. 
In order to estimate the number of detectable neutrino sources we introduce the monochromatic neutrino flux (in units of ${\rm GeV}\,{\rm cm}^{-2}\,{\rm s}^{-1}$) of a PS at redshift $z$ as:
\begin{equation}\label{eq:F}
F_{\nu} \equiv [E_\nu^2\phi_\nu^{\rm PS}]_{E_\nu=100\,{\rm TeV}} = \frac{(1+z)^{2-\gamma}}{4\pi d_L^2(z)}L_\nu\,.
\end{equation}
On average, we can expect that extragalactic neutrino sources distribute uniformly in the sky. The distribution of sources with flux $F_\nu$ and solid angle $\Omega$ is therefore:
\begin{equation}\label{eq:skyaveraged}
\frac{d^2N}{d F_\nu \, d\Omega} = \frac{1}{4\pi}  \int dz  \frac{dV_c}{dz}\frac{dL_\nu}{dF_\nu}\frac{d^2N}{dL_{\nu} dV_c}(L_{\nu}(F_\nu,z),z) \,.
 \end{equation}
The discovery potential (DP) of a time-integrated neutrino PS flux, $F_{\rm DP}$, typically has a strong dependence on the source declination $\delta$, which determines the level of atmospheric background of muons and neutrinos averaged over time. Using Eq.~(\ref{eq:skyaveraged}) we can account for the declination-dependence of the DP to estimate the number of expected discoveries as:
\begin{equation}\label{eq:N_DP}
N_{\rm DP} = 2 \pi \int_{-1}^{1} d \sin \delta \int_{F_{\rm DP}(\delta)}^{\infty} dF_\nu \, \frac{d^2N}{dF_\nu \, d\Omega}\,.
\end{equation}
In the following, we will use IceCube's DP of muon neutrinos via track-like events based on ten years of data~\cite{IceCube:2019cia}. Note that previous estimates have accounted for the declination dependence of the DP in an approximate way, {\it e.g.}~by restricting PS discoveries to the Northern Hemisphere with lower atmospheric background, {\it e.g.}~\cite{Ahlers:2018fkn,Ackermann:2019ows,Ackermann:2022rqc}. 

\begin{figure*}[t!]\centering
\includegraphics[width=0.5\linewidth]{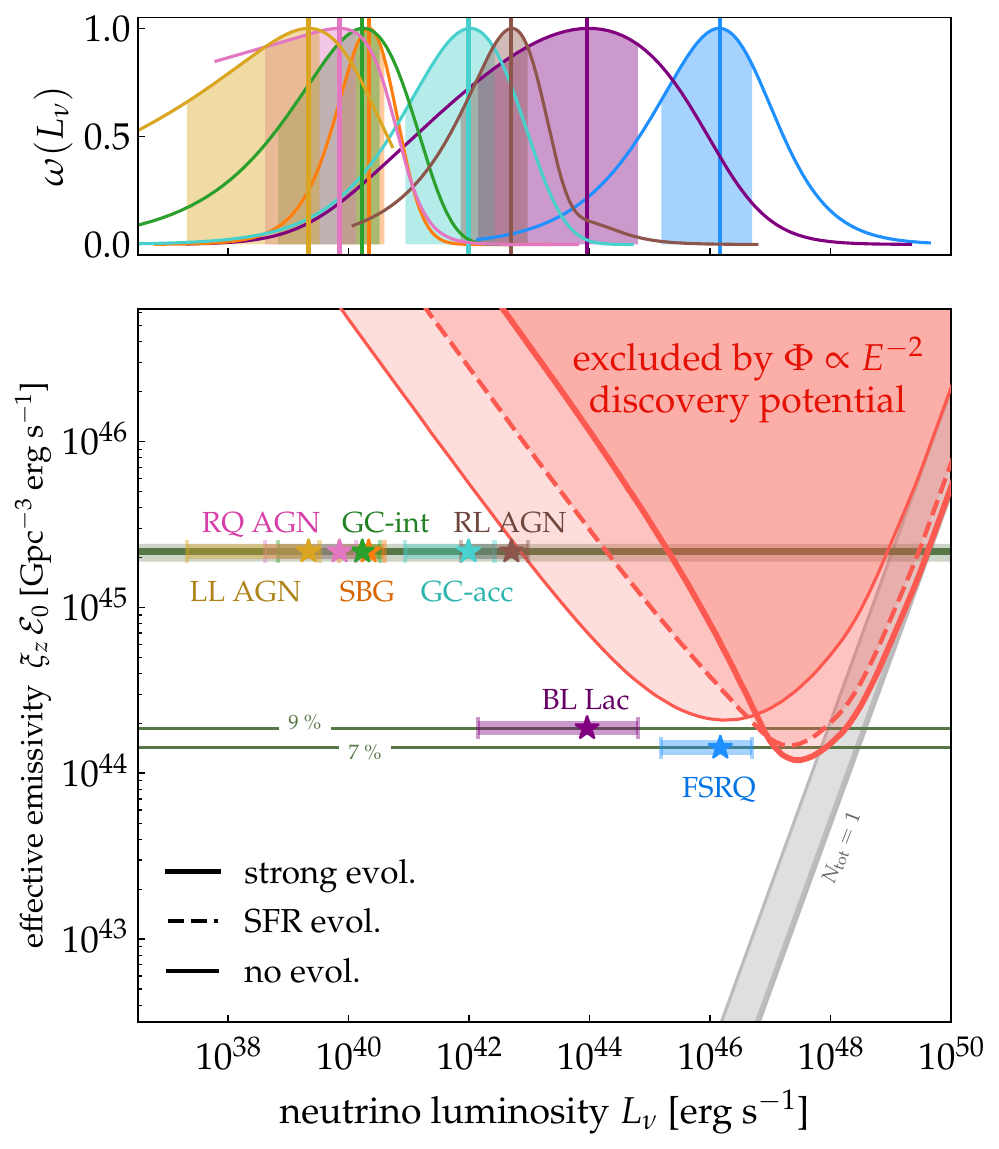}\hfill\includegraphics[width=0.5\linewidth]{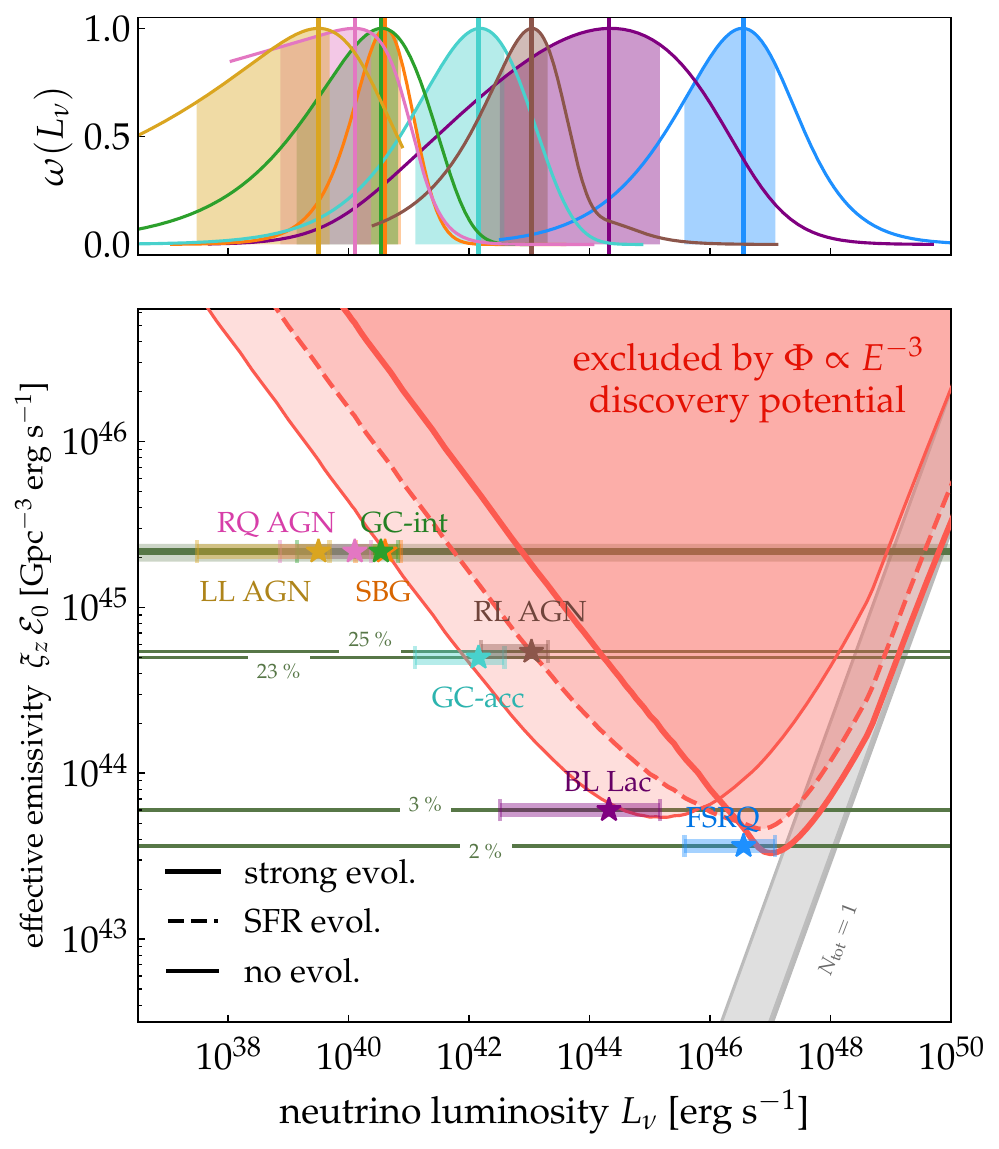}
\caption{\label{fig:rhoL} 
Constraints on extragalactic source populations in terms of their neutrino luminosity and effective emissivity based on IceCube's 10 yr point-source discovery potential~\cite{IceCube:2019cia} for spectral index $\gamma=2$ (left) and $\gamma=3$ (right). 
\textit{Top panels:} The local luminosity distribution of Eq.~(\ref{eq:w}) for each source class, with peak luminosity and 50\% central range indicated as vertical lines and shaded regions, respectively. 
\textit{Bottom panels:} 
The red filled contours show the excluded combinations of neutrino luminosity and emissivity of standard-candle populations assuming three cases of redshift evolution for the comoving number density (see Section~\ref{sec2}). The green horizontal band shows the effective emissivity of IceCube from Eq.~(\ref{eq:emissivityIC}). The stars and horizontal bands show the peak luminosity and 50\% central range for different source classes at the maximum emissivity consistent with the non-detection of the brightest source, $N_{\rm DP} = 1$, using Eq.~(\ref{eq:N_DP}).
See Section~\ref{sec3} and Table \ref{tab:Luminosity} for further details on the source candidates.}
\end{figure*}

Before we consider specific source populations characterized by their LF, we first discuss generic limits on the source populations that can be approximated by standard candles with fixed luminosity $L^\star_\nu$. In this case, the LF is separable as in Eq.~(\ref{eq:separable}) with $f^\star(L_\nu) = \delta(L_{\nu}-L_{\nu}^\star)$ and $\langle L_\nu\rangle = L_{\nu}^\star$. The number of expected PS discoveries is here simply:
\begin{equation}\label{eq:N_DPstar}
N^\star_{\rm DP} = \frac{1}{2} \int_{-1}^{1} d \sin \delta \int_0^{z^\star(\delta)} dz  \frac{dV_c}{dz} \rho(z)\,,      
\end{equation}
where $z^\star(\delta)$ is the redshift discovery horizon of a standard candle at declination $\delta$ following from $F_\nu(z^\star) = F_{\rm DP}(\delta)$. By setting $N^\star_{\rm DP}=1$, we can identify standard-candle populations with luminosity $L^\star_{\nu}$ and local density $\rho_0$ excluded by the non-discovery of PSs. Figure~\ref{fig:rhoL} shows these standard-candle exclusion contours in terms of the (evolution-corrected) local emissivity $\xi_z\mathcal{E}_0$ and (average) luminosity $L_\nu$ using the 5$\sigma$ point-source DP of Ref.~\cite{IceCube:2019cia} assuming a spectral index $\gamma=2$ (left plot) and $\gamma=3$ (right plot). The figure compares the exclusion contours with the requirements from observations of the diffuse astrophysical neutrino flux. We also show various candidate neutrino source classes discussed in Section~\ref{sec3}. 

The exclusion contours in Fig.~\ref{fig:rhoL} exhibit a strong dependence on the redshift evolution of the sources, which is indicated for the three benchmark cases of no evolution (thin red line), moderate evolution similar to the SFR (dashed red line) or strong evolution (bold red line). Populations with a weak redshift evolution and, correspondingly, a low $\xi_z$ will require a larger local luminosity density $\mathcal{E}_0$ to reproduce the same diffuse flux level. For a fixed luminosity, this in turn requires a larger local density $\rho_0$ with brighter local sources. We find that standard candle populations with an $E^{-2}$ emission spectrum (left plot) cannot contribute with 100\% of the observed diffuse neutrino flux for luminosities $L^\star_\nu \gtrsim 7\times 10^{42}$~erg/s ($L^\star_\nu \gtrsim 2\times10^{45}$~erg/s) and local densities $\rho_0 \lesssim 600 \, \rm{Gpc^{-3}}$ ($\rho_0 \lesssim 0.09 \, \rm{Gpc^{-3}}$) assuming no (strong) redshift evolution. 

The overall luminosity scaling of the exclusion contours in Fig.~\ref{fig:rhoL} can be understood in terms of the required local density $\rho_0$ at a fixed emissivity level. For high $\rho_0$, {\it i.e.}~towards the top left of the plots, the brightest sources appear nearby ($z\ll 1$). In this case, the luminosity distance in Eq.~(\ref{eq:PS}) is linear in redshift $d_L \simeq r = zR_H$ and the flux distribution of sources follows that expected from Euclidean space. From $F_\nu \simeq L^\star_\nu/(4\pi r^2)$ and $dN/dr \simeq \rho_0 4\pi r^2$ we can derive the expected flux distribution as:
\begin{equation}\label{eq:dNdF}
\frac{dN}{dF_\nu} \simeq \frac{3}{2}\frac{1}{F_0}\left(\frac{F_\nu}{F_0}\right)^{-5/2}\,,
\end{equation}
where we introduce the flux $F_0 = L^\star_\nu/(4\pi r_0^2)$ for a source located at a distance $r_0$ that defines a sphere containing one source, $4\pi\rho_0r_0^3/3 = 1$. The total number of sources with a flux larger than $F_\nu$ is then simply:
\begin{equation}\label{eq:NF}
N(F_\nu) \simeq \left(\frac{F_\nu}{F_0}\right)^{-3/2}\,,
\end{equation}
and the expected number of observed sources in Eq.~(\ref{eq:N_DP}) scales then as:
\begin{equation}\label{eq:N_DPstar2}
N^\star_{\rm DP} \simeq \frac{1}{2}\int_{-1}^{1} d \sin \delta \left(\frac{F_{\rm DP}(\delta)}{F_0}\right)^{-3/2}\,.    
\end{equation}
For $N^\star_{\rm DP}=1$, we therefore obtain an upper limit on the local density that decreases with luminosity as $\rho_0 \propto (L_\nu^\star)^{-3/2}$ and the emissivity in Fig.~\ref{fig:rhoL} is limited as $\xi_z\mathcal{E}_0 \propto (L_\nu^\star)^{-1/2}$.

On the other hand, for very low local densities, $\rho_0 \lesssim 10\,{\rm Gpc}^{-3}$, {\it i.e.}~towards the bottom right of the plots in Fig.~\ref{fig:rhoL}, the brightest sources appear at high redshift and the Euclidean approximation fails. This becomes noticeable in the exclusion contours of Fig.~\ref{fig:rhoL} already for luminosities $L_\nu^\star \gtrsim 10^{44} {\rm erg}/{\rm s}$. Ultimately, for further decreasing source densities the total number of sources:
\begin{equation}\label{eq:Ntot}
N_{\rm tot} = \int dz\frac{dV_c}{dz}\rho(z)\,,
\end{equation}
becomes $\mathcal{O}(1)$ and a discussion in terms of source populations becomes inadequate. The limit $N_{\rm tot} =1$ is indicated in Fig.~\ref{fig:rhoL} as the band, from no evolution (thin grey line) to strong evolution (bold grey line). The exclusion contours are seen to be well bounded by their corresponding $N_{\rm tot} =1$ limits for low densities/high luminosities.

\begin{figure}[t!]
\includegraphics[width=\linewidth]{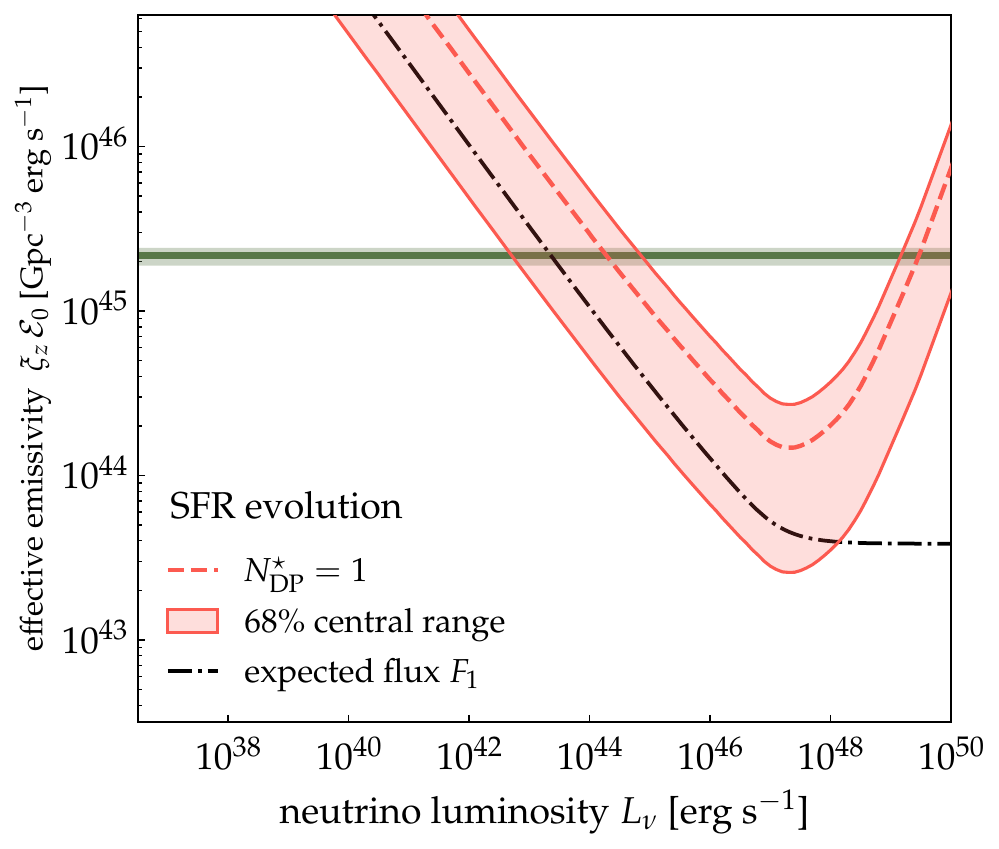}
\caption{\label{fig:variance} 
Effect of ensemble fluctuations on the exclusion contours illustrated for the case of standard candles following the SFR evolution. The green horizontal band shows the effective emissivity of IceCube from Eq.~(\ref{eq:emissivityIC}). The thin red lines outline the 68\% central region accounting for Poisson fluctuations of the expectation value $N^\star_{\rm DP}$. The dashed red line shows $N^\star_{\rm DP}=1$. The dash-dotted black line shows the exclusion contour based on the expected flux of the closest source. See Section \ref{sec2} for details.}
\end{figure}

Note that our limits in Fig.~\ref{fig:rhoL} appear to be weaker compared to earlier treatments based on IceCube's discovery potential, {\it e.g.}~\cite{Ackermann:2019ows,Gen2_TDR}; see also~\cite{Murase:2016gly}. For instance, Ref.~\cite{Gen2_TDR} argues that IceCube presently excludes non-evolving standard candles at the level of $L^\star_\nu \gtrsim 2\times10^{41} \, {\rm erg}/{\rm s}$. These earlier treatments derived exclusion limits based on the expected flux $F_1$ of the closest source of the population. For a flux distribution in Euclidean space (\ref{eq:dNdF}) we expect that the closest source will appear with an average flux $F_1 \simeq 2.7 F_0$~\cite{Ahlers:2018fkn}; see also Section~\ref{sec4}. This small increase in flux has a significant impact on the exclusion limits. In Euclidean space, the flux of the brightest ({\it i.e.}~closest) sources scales as $F_\nu \propto L_\nu \rho_0^{2/3}$. If the local emissivity $\mathcal{E}_0 = \rho_0L^\star_\nu$ is fixed to IceCube's diffuse flux the dependence becomes $F_\nu \propto L_\nu^{1/3}$. The limit on the luminosity therefore decreases by a factor $(F_1/F_0)^3 \simeq 20$ if one were to compare to $F_1$.

We argue that our prescription $N^\star_{\rm DP}=1$ based on Eqs.~(\ref{eq:N_DPstar2}) is a more robust way to estimate the present reach of neutrino telescopes to constrain source populations. Firstly, while Eq.~(\ref{eq:N_DPstar2}) allows us to account for the full declination-dependence of IceCube's DP, the exclusion limits for based on the mean flux $F_1$ only do this in an approximate way, by comparing $F_1$ to the {\it average} DP in the Northern Hemisphere as $\langle F_{\rm DP}\rangle_{\rm North} = f_{\rm sky}^{2/3}F_1$ with $f_{\rm sky} = 1/2$; see Ref.~\cite{Ahlers:2018fkn} for details. And, secondly, it is only in 20\% of the cases that a source appears with a flux $F\geq F_1$ while it is expected in 63\% of the cases for $F\geq F_0$. This shows that ensemble fluctuations related to the distribution of sources can have a significant impact on the discovery of individual bright sources of the population. The probability that at least one source of a source population has been discovered is given by $P_{\rm DP} = 1-e^{-N_{\rm DP}}$ using Eq.~(\ref{eq:N_DPstar}). We can estimate the variance of the exclusion limits by showing $N_{\rm DP} \simeq 0.17$ and $N_{\rm DP} \simeq 1.83$, corresponding to the 68\% central range of the $N_{\rm DP}$-distribution $dP_{\rm DP}/dN_{\rm DP}$. 

Figure~\ref{fig:variance} shows the effect of ensemble fluctuations on the exclusion contours assuming redshift evolution following the SFR. The thin red lines show the 68\% central region of ensemble variations, whereas the dashed red line indicates the contour $N^\star_{\rm DP}=1$ from Fig.~\ref{fig:rhoL}. We see that ensemble fluctuations have a strong effect on the luminosity requirements for the discovery of local sources, introducing a broad window of $\Delta L_\nu /L_\nu \simeq (1.83/0.17)^2 \simeq 116$ for a fixed local emissivity. The dash-dotted black line indicates the exclusion limit from the expected flux $F_1$ of the brightest source in comparison to $\langle F_{\rm DP}\rangle_{\rm North} \simeq 2\times10^{-12}\,{\rm TeV}\,{\rm cm}^{-2}\,{\rm s}^{-1}$ (per flavour) from Ref.~\cite{IceCube:2019cia}. This bound is consistent with the 68\% central range for the brightest source. Note that the line continues into the region to the right of the plot where the source numbers become too low to meaningfully talk about populations of sources. This adds to the advantages of using the expected number of sources over using the expected flux as the discovery threshold.

\begin{table*}[t!]
    \caption{\label{tab:Luminosity}
    Characteristics of neutrino and photon luminosity distributions for different source types.
    All neutrino luminosities are summed over neutrino flavours, for both neutrinos and anti-neutrinos.
    }
    \centering\renewcommand{\arraystretch}{1.6}
    \begin{ruledtabular}
    \begin{tabular}{cccccccccc}
    Population & 
      \renewcommand{\thempfootnote}{\fnsymbol{mpfootnote}}
    $L_{\nu}(L_{\rm ph})$\footnotemark[1] &
      \renewcommand{\thempfootnote}{\fnsymbol{mpfootnote}}
	  $L_{\mathrm{ph}}^{\rm pk}$\footnotemark[1] & 
      \renewcommand{\thempfootnote}{\fnsymbol{mpfootnote}}
	  $L_{\nu}^{\rm pk}$\footnotemark[1] &
      \renewcommand{\thempfootnote}{\fnsymbol{mpfootnote}}
	  $L_{\nu}^{50\%}/L_{\nu}^{\rm pk}$\footnotemark[1] &
      \renewcommand{\thempfootnote}{\fnsymbol{mpfootnote}}
	  $\rho_0^{\rm pk}$ \footnotemark[1] & 
      \renewcommand{\thempfootnote}{\fnsymbol{mpfootnote}}
	  $\xi_z$\footnotemark[2]  &
    $\xi_z^{\mathrm{pk}}$  &
    $N_{\mathrm{DP}}$ &
    Refs.  \\[-0.1cm]
     &&[erg/s]&[erg/s]& &[Gpc${}^{-3}$]& $\gamma=2$ (${3}$)   & $\gamma=2$ (${3}$)  & $\gamma=2$ (${3}$) & \\[0.1cm]
    \colrule 
  FSRQ &
    $L_{\nu}\propto L_{\gamma}^{3/2}$  &
    $5\times 10^{47}$ &  
	  $1 \times 10^{46}$ &
	  $0.1 - 3 $  &
	  $5 \times 10^{-3} $  & 
    $31.8$ ($13.1$) &
    $20.7$ ($10.2$) &
	  $15$ ($60$) &
    \cite{Ajello:2011zi, Murase:2014foa} \\
	\colrule
  BL~Lac &
    $L_{\nu}\propto L_{\gamma}^{2}$  &
    $2\times 10^{46}$ & 
	  $9 \times 10^{43}$ &
	  $0.02 - 7$ &
	  $6 \times 10^{-1}$ &
    $39.3$ ($17.0$) &
    $1.9$ ($1.2$) &
	  $12$ ($36$) &
    \cite{Ajello:2013lka, Tavecchio:2014eia} \\
	\colrule
  SBG &
    $L_{\nu}\propto L_{\gamma}\propto L_{\rm IR}^{1.17}$  &
    $1\times 10^{41}$ & 
    $2 \times 10^{40}$ &
    $0.3 - 1.8 $  &
    $3 \times 10^{4} $ & 
    $2.9$ ($1.6$) &
    $2.9$ ($1.6$) &
    $0.01$ ($0.3$) &
    \cite{Gruppioni:2013jna,Fermi-LAT:2012nqz,Loeb:2006tw}\\
    \colrule 
  GC-acc &
    $L_{\nu}\propto M^2 \propto L_{X}^{3}$  &
    $3\times 10^{44}\,$ &
    $1 \times 10^{42}$ &
    $0.09 - 3$  &
    $4 \times 10^{3} $   &  
    $0.5$ ($0.4$) &
    $0.5$ ($0.4$) &
    $0.3$ ($4$) &
    \cite{Warren:2005ey,Reiprich:2001zv,Diemer:2017bwl,Zandanel:2014pva,Murase:2016gly} \\
    \colrule 
  GC-int &
    $L_{\nu}\propto M^{4/3} \propto L_{X}^{2}$ &
    $5\times 10^{43}\,$ & 
    $2 \times 10^{40}$ &
    $0.04 - 1.9 $  &
    $5 \times 10^{4} $  & 
    $2.7$ ($1.3$) &
    $2.7$ ($1.3$) &
    $0.01$ ($0.2$) & 
    \cite{Warren:2005ey,Reiprich:2001zv,Diemer:2017bwl} \\
    \colrule 
  RL AGN & 
    $L_{\nu}\propto L_{\gamma} \propto L_{\rm radio}^{1.16}$ &
    $2 \times 10^{45}$ & 
    $5 \times 10^{42}$ &
    $0.1 - 1.9$  &
    $9 \times 10^{1} $  & 
    $4.6$ ($2.2$) &
    $2.0$ ($1.1$) &
    $0.2$ ($4$) &
    \cite{Willott:2000dh,Inoue:2011bm,Murase:2016gly}
    \\
    \colrule 
  RQ AGN &
    $L_{\nu}\propto L_{X}$ & 
    $1 \times 10^{43}\,$ & 
    $7 \times 10^{39}$ &
    $0.06-1.9$  &
    $1 \times 10^{5} $  & 
    $2.1$ ($1.2$) &
    $1.8$ ($1.1$) &
    $0.01$ ($0.2$) & 
    \cite{Ueda:2014tma,Murase:2016gly} \\
    \colrule 
  LL AGN  &
    $L_{\nu}\propto L_{\rm{H}{\alpha}}$ &
    $4 \times 10^{40}\,$ & 
    $2 \times 10^{39}$ &
    $0.01 - 1.5 $  &
    $2 \times 10^{6} $  & 
    $0.5$ ($0.4$) & 
    $0.5$ ($0.4$) & 
    $0.02$ ($0.3$) &
    \cite{Ho:2008rf, SDSS:2005lxf, Murase:2016gly} \\
    \end{tabular}
    \end{ruledtabular}
  \renewcommand{\thempfootnote}{\fnsymbol{mpfootnote}}
  \footnotetext[1]{These values use spectral index $\gamma=2$.}
  \footnotetext[2]{Based on Eq.~(\ref{eq:xi_F}) assuming a neutrino luminosity function derived from Eqs.~(\ref{eq:Lnu}) and (\ref{eq:dNdL}).}
\end{table*}

\section{\label{sec3} Candidate Neutrino Sources}

So far, our discussion has focused on generic source populations that can be approximated as standard candles. We will now turn to specific populations that have been considered as candidate sources of steady high-energy neutrino emission that can reproduce IceCube's diffuse flux. The neutrino LFs of potential sources are generally not well known. However, in multi-messenger sources we can relate the neutrino luminosity to that of cosmic rays (CRs) and photons. Our discussion will closely follow the methodology of {\it Murase \& Waxman}~\cite{Murase:2016gly} who introduced a recipe to extract the neutrino luminosity distribution from scaling relations to photon luminosities.  

High-energy neutrino emission is the result of hadronic interactions of high-energy cosmic rays with gas ($pp$) and radiation ($p\gamma$). Charged pions produced in these interactions decay via $\pi^- \to \mu^-+\bar\nu_\mu$ followed by $\mu^- \to e^-+\bar\nu_e+\nu_\mu$ and the charge-conjugate processes. Typically, pions carry about 1/5 of the initial energy of CR nucleons and each of the three neutrinos receives about 1/4 of the pion energy. The neutrino luminosity $L_\nu$ at 100~TeV summed over flavours is therefore related to the luminosity $L_{\rm CR}$ of CR nucleons at about 2~PeV as~\cite{Ahlers:2018fkn}:
\begin{equation}\label{eq:Lnu1}
L_\nu \simeq \frac{3}{4}(1-e^{-\kappa\tau})\frac{K_\pi}{1+K_\pi}L_{\rm CR}\,,
\end{equation}
where $K_\pi\simeq 1$ ($K_\pi\simeq 2$) is the ratio of charged-to-neutral pions produced in $p\gamma$ ($pp$) interactions with inelasticity $\kappa\simeq 0.2$ ($\kappa\simeq 0.5$) and $\tau$ the opacity of the source environment. For opaque sources ($\tau \gg 1$) we expect that the neutrino luminosity is proportional to that of CRs, $L_\nu \propto L_{\rm CR}$, whereas transparent sources ($\tau \ll 1$) can introduce an additional luminosity dependence via the opacity, $L_\nu \propto \tau L_{\rm CR}$. 

In parallel, neutral pions from CR interactions decay as $\pi^0 \to \gamma+\gamma$ producing $\gamma$-rays with about twice the energy of the corresponding neutrinos. The neutrino luminosity $L_\nu$ can therefore be related to the that of hadronic $\gamma$-rays with energies of about 200~TeV as~\cite{Ahlers:2018fkn}:
\begin{equation}\label{eq:Lnu2}
L_\nu \simeq \frac{3}{4}K_\pi  L_{\gamma}\,.
\end{equation}
After correcting for $\gamma$-ray absorption in the source and during propagation, this also motivates a linear dependence, $L_\nu \propto L_\gamma$, assuming $\gamma$-rays as a proxy.

In multi-messenger sources, the CR luminosity can in general be related to that of some proxy electromagnetic emission, $L_{\rm CR} \propto L_{\rm ph}$. For opaque CR sources or neutrino sources dominated by $pp$ interactions, we arrive at $L_\nu \propto L_{\rm ph}$. For transparent CR sources dominated by $p\gamma$ interactions, {\it e.g.}~blazars, the optical depth depends on the density of target photons. Efficient pion production via the $\Delta$ resonance requires target photons with energy $\varepsilon_t \simeq 8(\Gamma/10)^2(E_\nu/100\,{\rm TeV})^{-1}{\rm \, keV}$, where $\Gamma$ is the bulk Lorentz factor of the source environment~\cite{Murase:2015xka}. The relevant opacity for 100~TeV neutrino production depends on the luminosity of target photons observed as X-rays or soft $\gamma$-rays, $\tau \propto L_{\rm t}$. Assuming equipartition in steady sources, $L_{\rm t} \propto L_{\rm CR} \propto L_{\rm ph}$, this results in the dependence $L_{\rm CR} \propto L^2_{\rm ph}$.

The previous two scaling relations do not reproduce all possibilities. Following Ref.~\cite{Murase:2015xka}, we assume a power-law relation between the photon luminosity and neutrino luminosity of the form:
\begin{equation}\label{eq:Lnu}
L_{\nu} = L_{\nu,0} \left(\frac{L_{\mathrm{ph}}}{L_{\mathrm{ph},0}}\right)^{\alpha}\,, 
\end{equation}
where the scaling power $\alpha$ depends on the neutrino production mechanism and its relation to the luminosity $L_{\rm ph}$ of electromagnetic emission used as a proxy.
The reference luminosities $L_{\nu,0}$ and $L_{\mathrm{ph},0}$ appearing in Eq.~(\ref{eq:Lnu}) depend on details of the source environment. However, assuming that the source population is responsible for the diffuse flux observed by IceCube, we can fix the normalization $L_{\nu,0}/L_{\mathrm{ph},0}^\alpha$ by Eqs.~(\ref{eq:E2Phi}), (\ref{eq:emissivity}) and (\ref{eq:E2PhiIC}). This also fixes the neutrino LF as:
\begin{equation}\label{eq:dNdL}
\frac{d^2N}{dL_\nu dV_c} = \frac{1}{\alpha}\frac{L_\mathrm{ph}}{L_\nu}\frac{d^2N}{dL_\mathrm{ph} dV_c}\,,
\end{equation}
and allows us to determine the expected number of observable neutrino sources of that population via Eqs.~(\ref{eq:skyaveraged}) and (\ref{eq:N_DP}).

Following Ref.~\cite{Murase:2016gly}, we consider a collection of relevant candidate neutrino source types, listed in Table~\ref{tab:Luminosity}: 
\begin{itemize}[leftmargin=*]\setlength\itemsep{0pt}
\item Flat-Spectrum Radio Quasars (FSRQ):
We assume CR interactions with radiation from the broadline region and dust torus, which is reprocessed emission from the accretion disk of the central supermassive black hole~\cite{Murase:2014foa}. The opacity of the optically thin regions scales with the luminosity of the disk as $\tau \propto L^{1/2}_{\rm disk}$~\cite{Ghisellini:2008zp}. Using $\gamma$-ray luminosity as a proxy, the neutrino luminosity becomes $L_\nu \propto L_\gamma^{3/2}$. To model the $\gamma$-ray luminosity of FSRQs from $10^{45}\,{\rm erg}\,{\rm s}^{-1}$ to $10^{50}\,{\rm erg}\,{\rm s}^{-1}$, we used the luminosity-dependent density evolution (LDDE) ``ALL" best-fit LF model and redshift evolution of FSRQs in \cite{Ajello:2011zi}.
\item BL~Lac objects: 
We consider CR interactions with synchrotron photons in the jet~\cite{Murase:2014foa} and assume $\tau \propto L_{\rm syn}$ following the model of Ref.~\cite{Tavecchio:2014eia}. Again, assuming $\gamma$-rays as a proxy for CR and jet emission, the neutrino luminosity is expected to scale as $L_\nu \propto L_\gamma^{2}$ in an optically thin jet. We use the LDDE${}_2$ best-fit model and redshift evolution of \cite{Ajello:2013lka} for the BL Lac $\gamma$-ray luminosity function for $L_\gamma$ from $10^{43}{\rm erg}\,{\rm s}^{-1}$ to  $10^{49}\,{\rm erg}\,{\rm s}^{-1}$.
\item Starburst Galaxies (SBG):
Neutrinos and $\gamma$-rays are produced in charged and neutral pion production from CR interactions in the starburst interstellar medium, whereby $L_{\nu} \propto L_{\gamma}$ is expected~\cite{Loeb:2006tw,Tamborra:2014xia,Peretti:2019vsj}. As a proxy, we use infrared (IR) emission with LF and redshift evolution of SBGs of Ref.~\cite{Gruppioni:2013jna} and the relation $L_{\gamma} \propto L^{1.17}_{\rm IR}$ found in Ref.~\cite{Fermi-LAT:2012nqz}, for $L_\gamma$ from $3\times 10^{37}{\rm erg}\,{\rm s}^{-1}$ to  $10^{44}\,{\rm erg}\,{\rm s}^{-1}$.
\item Galaxy Clusters (GC) with CR acceleration in accretion shocks (GC-acc):
For a cluster halo mass $M$ and accretion rate $\dot M$, accretion shocks are expected to form at a virial radius $r\propto M^{1/3}$ leading to a CR luminosity scaling of $L_{\rm CR} \propto M\dot M/r \propto M^{5/3}$~\cite{Keshet:2004dr,Murase:2013rfa}. CR diffusion and interactions with gas in the GC environment lead to $\tau \propto r^2 \propto M^{2/3}$. This leads to $L_{\nu} \propto M^{5/3}$ ($\tau\gg1$) or $M^{7/3}$ ($\tau\ll1$) and we use the intermediate relation $L_{\nu} \propto M^2$ following Ref.~\cite{Murase:2016gly}. The cluster halo mass is inferred from $X$-ray emission using the $M\propto L^{1/5}_{\rm X}$ following Ref.~\cite{Reiprich:2001zv}, for $L_{X}$ from $2\times 10^{37}\,{\rm erg}\,{\rm s}^{-1}$ to $3\times 10^{46}\,{\rm erg}\,{\rm s}^{-1}$. We use the halo mass function of Ref.~\cite{Warren:2005ey,Diemer:2017bwl} and assume no redshift evolution (see {\it e.g.}~Ref.~\cite{Zandanel:2014pva}). 
\item Galaxy Clusters with CR acceleration in sources (GC-int):
In this model, the CR production happens in sources hosted by galaxy clusters (or groups) and the CR luminosity is assumed to be proportional to the halo mass, $L_{\rm CR} \propto M,$ with a SFR-like redshift evolution. Assuming CR interactions in the GC environment as in the case of GC-acc, we either have $L_{\nu} \propto M$ ($\tau\gg1$) or $M^{5/3}$ ($\tau\ll1$) and we settle on $L_{\nu} \propto M^{4/3}$. The same halo mass function and relation to X-ray luminosity is used as in the GC-acc case above. 
\item Misaligned Radio-Loud (RL) AGN:
Cosmic rays that are accelerated in (misaligned) jets can escape into the surrounding medium. Assuming that $\gamma$-ray emission is dominated by pion production from $pp$ interactions~\cite{Murase:2013rfa}, we expect $L_\nu \propto L_\gamma$. Combining this with relation of $\gamma$-ray luminosity to radio luminosity~\cite{Inoue:2011bm} we arrive at $L_\nu \propto L^{1.16}_{\rm radio}$. For modelling the radio luminosity function, we use the LDDE \textit{Model C 1} and redshift evolution of \cite{Willott:2000dh} for $L_\gamma$ from $4\times 10^{42}\,{\rm erg}\,{\rm s}^{-1}$ to $2\times 10^{49}\,{\rm erg}\,{\rm s}^{-1}$.
\item Radio-Quiet (RQ) AGN:
For these types of AGN without strong jets, CRs can be accelerated in accretion shocks and in turbulence of accretion flows or the corona in the vicinity of the supermassive black hole and subsequently interact with the radiation background~\cite{Stecker:1991vm,Alvarez-Muniz:2004xlu}. Assuming $p\gamma$ interactions with $X$-ray emission of the AGN core leads to $L_\nu \propto L_X$~\cite{Murase:2015xka,Murase:2019vdl,Inoue:2019fil,Kheirandish:2021wkm}. We model the X-ray LF with the LDDE model and (strong) redshift evolution of Ref.~\cite{Ueda:2014tma} for $L_X$ from $10^{41}\,{\rm erg}\,{\rm s}^{-1}$ to $10^{47}\,{\rm erg}\,{\rm s}^{-1}$.
\item Low-Luminosity (LL) AGN:
Neutrinos are assumed to be primarily produced in $pp$ interactions in radiatively inefficient accretion flows near the central black hole~\cite{Kimura:2014jba}. Following Ref.~\cite{Murase:2016gly}, the proxy luminosity for that of CRs is here the strength of the $\rm{H}{\alpha}$ line emission leading to $L_\nu \propto L_{\rm{H}{\alpha}}$. We use the $\rm{H}{\alpha}$ LF of Refs.~\cite{Ho:2008rf, SDSS:2005lxf} for $L_{\rm{H}\alpha}$ from $6\times 10^{36}\,{\rm erg}\,{\rm s}^{-1}$ to $10^{42}\,{\rm erg}\,{\rm s}^{-1}$ and assume no redshift evolution as in Ref.~\cite{Murase:2016gly}.
\end{itemize}
Each of the source candidates listed above is characterised by a scaling power $\alpha$ of the neutrino luminosity to a proxy photon luminosity and the corresponding photon LF. Given the corresponding neutrino LF in Eq.~(\ref{eq:dNdL}), we can calculate the expected number of discovered PSs through Eqs.~(\ref{eq:skyaveraged}) and (\ref{eq:N_DP}). The results are listed in the last column of Table~\ref{tab:Luminosity} for two assumptions of the source spectral index $\gamma$. 
A value $N_{\rm DP} > 1$ indicates that the population cannot account for 100\% of the diffuse flux.

While the calculation of $N_{\rm DP}$ for the source candidates does not require a specific form of the neutrino LF, it is instructive to compare these results to our previous exclusion limits based on standard-candle populations shown in the lower panels of Fig.~\ref{fig:rhoL}. We expect that these generic exclusion limits provide a good approximation if the neutrino LF appearing in the emissivity of Eq.~(\ref{eq:emissivity}) is sufficiently narrow and centered on a peak luminosity $L_\nu^{\rm pk}$ that has a weak dependence on redshift. The top panels in Fig.~\ref{fig:rhoL} show the relative distribution of neutrino luminosities weighted by the neutrino LF (\ref{eq:dNdL}) at redshift $z=0$ normalized to $L_\nu^{\rm pk}$ as:
\begin{equation} \label{eq:w}
\omega(L_\nu) \equiv L^2_\nu\frac{d^2N_0}{dL_\nu dV_c}\bigg/\left[L^2_\nu\frac{d^2N_0}{dL_\nu dV_c}\right]_{L_\nu = L^{\rm pk}_\nu}\,.
\end{equation}
For some source candidates, the distribution is narrow and well represented by the peak luminosity, whereas for others, such as LL~AGN or BL~Lac, the distribution is very broad. In comparing to generic populations of standard candles in the lower panel, we indicate the neutrino luminosity distribution of the populations by their peak luminosity $L_\nu^{\rm pk}$ (star symbols) and the $50$\% central range (horizontal shaded bands). These values are also listed in Table~\ref{tab:Luminosity} together with the required local source density $\rho_0^{\rm pk}$ to reach 100\% of the diffuse flux, inferred from the peak neutrino luminosity.

\begin{figure}[t]
 \includegraphics[width=\linewidth]{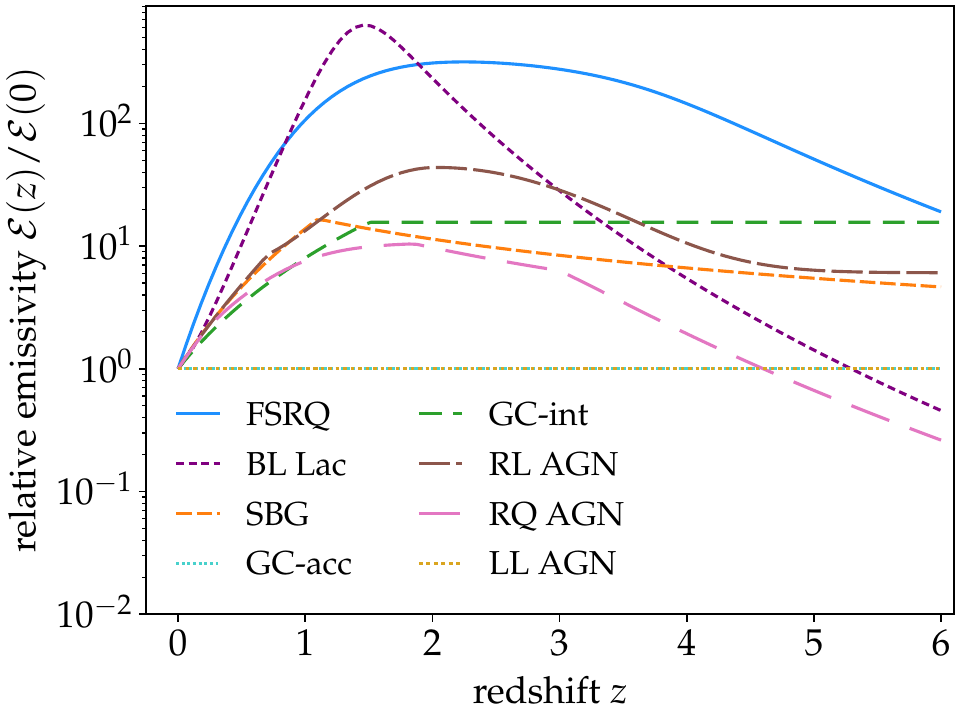}
\caption{\label{fig:Ez}
Relative emissivity given by Eq.~(\ref{eq:emissivity}) for the different source models listed in Table~\ref{tab:Luminosity}, normalized to the emissivity at redshift $z=0$.}
\end{figure}

In general, we observe that, consistent with our earlier discussion, populations with relatively low peak luminosity can account for 100\% of the observed neutrino flux, corresponding to the effective emissivity level~(\ref{eq:emissivityIC}) shown as the horizontal green line. For source classes where the expected number of observed sources is greater than 1 (see Table~\ref{tab:Luminosity}), the contribution to the diffuse flux is limited to a fraction of $1/N_{\rm DP}$, indicated as thin green lines. The populations where the neutrino LF in Eq.~(\ref{eq:dNdL}) peaks at very high luminosities, FSRQs and BL~Lacs, are each ruled out as the dominant source class for the diffuse flux assuming the neutrino luminosity scaling indicated in the second column of Table~\ref{tab:Luminosity}. Assuming a spectral index of $\gamma=2$ (left plot of Fig.~\ref{fig:rhoL}), we find $N_{\mathrm{DP}} \simeq 15$ ($N_{\mathrm{DP}} \simeq 12$) for FSRQs (BL~Lacs), leading to upper limits of 7\% (9\%) for their maximal expected contribution to the observed diffuse flux. These results are consistent with blazar stacking limits set by IceCube~\cite{IceCube:2016qvd,Glusenkamp:2015jca}, with many other works also finding blazars to be subdominant with contributions to the diffuse flux of 1\%--10\%, {\it e.g.}~\cite{Murase:2018iyl,Bartos:2021tok}. For a softer spectrum with $\gamma=3$ (right plot) the exclusion limits for FSRQs (BL~Lacs) become stronger with $N_{\mathrm{DP}}\simeq 60$ ($N_{\mathrm{DP}}\simeq 36$) leading to an upper limit of about 2\% (3\%). In addition, the softer spectrum disfavours a wider range of source classes, as both RL AGN and GC-acc fall within the excluded relevant contours for SFR-like and weak evolution, respectively, at the 100\% level. The resulting numbers of expected sources are $N_{\mathrm{DP}}\simeq4$ for both RL AGN and GC-acc, corresponding to flux limits of 25\% and 23\%, respectively. 

The standard-candle exclusion contours in Fig.~\ref{fig:rhoL} represent three different redshift evolution cases, that are characterized by the evolution parameter $\xi_z$ with increasing value from weak to strong redshift evolution. The relative emissivity $\mathcal{E}(z)/\mathcal{E}(0)$ appearing in its definition in Eq.~(\ref{eq:xi_F}) is shown in Fig.~\ref{fig:Ez} for the individual source classes and the resulting evolution factor is listed in column 7 of Table~\ref{tab:Luminosity}. Regarding the limits of FSRQs (strong evolution), RL-AGN (intermediate evolution), and GC-acc (weak evolution), we find good agreement with the generic exclusion limits at the location of the peak luminosity in Fig.~\ref{fig:rhoL}. This shows that the three generic exclusion contours for standard candle populations, designed to represent the range of relevant evolution scenarios, are a good first-order approximation for these source classes.

An exception is the case of BL~Lacs. Here, the evolution parameter $\xi_z = 39.3$ $(17.0)$ from Eq.~(\ref{eq:xi_F}) with $\gamma = 2$ $(3)$ suggests a strong evolution and seems to disagree with the level of the corresponding exclusion limits in Fig.~\ref{fig:rhoL}. This shows that the case of BL~Lacs, which we model using the LDDE $\gamma$-ray LF of Ref.~\cite{Ajello:2013lka}, is not well described by a standard candle scenario. In fact, the neutrino luminosity distribution following from Eq.~(\ref{eq:dNdL}) is broad, as shown in the top panel of Fig.~\ref{fig:rhoL}, and we also find that the peak luminosity $L_\nu^{\rm pk}$ increases strongly with redshift. To illustrate this mismatch, we also indicate the redshift evolution factor $\xi_z^{\rm pk}$ derived from the density evolution fixed at $L_\nu^{\rm pk}(z=0)$ using the formulation of Ref.~\cite{Ahlers:2018fkn}. Whereas this factor would agree with $\xi_z$ for standard-candle populations, its value for BL~Lacs is significantly reduced.

In summary, the generic standard-candle exclusion contours in Fig.~\ref{fig:rhoL} provide meaningful first-order constraints on candidate populations as a dominant source of IceCube's diffuse flux, as long as their neutrino luminosity distributions are sufficiently narrow and their peak luminosities have a weak dependence on redshift. We also note that, for photon luminosity distributions with widths like the ones in Fig.~\ref{fig:rhoL}, the peak neutrino luminosity can have a strong dependence on the scaling power index $\alpha$ in Eq.~(\ref{eq:Lnu}). In any case, the expected number of observed PSs $N_{\rm DP}$ via Eq.~(\ref{eq:N_DP}) allows one to derive the flux fraction $1/N_{\rm DP}$ for arbitrary candidate sources.

\section{\label{sec4} Population Analyses}

Our discussion of extragalactic neutrino source populations has so far centered on constraints imposed by the non-discovery of the brightest source following from a $dN/dF_\nu$ distribution. We investigate in this section how catalogue searches would be able to improve these constraints, assuming that the $N_{\rm cat}$ catalogue members comprise the brightest sources of the population. The following two subsections discuss the impact of two idealized analyses: {\it a)} a weighted stacking analysis, where the relative weights of PSs are known, and {\it b)} an analysis of non-Poissonian fluctuations in event-count histograms, where relative weights are unknown.

In both cases, we consider populations with relatively high local densities $\rho_0 \gtrsim 10 \, {\rm Gpc}^{-3}$ so that the total number of sources defined via Eq.~(\ref{eq:Ntot}) is sufficiently large $N_{\rm tot} \gg 1$. In this situation, the flux $F_\nu$ expected from the $k$th brightest source follows the probability distribution:
\begin{equation}\label{eq:p_k}
p_k(F_\nu) = \frac{dN}{dF_\nu} \ \frac{\left(N(F_\nu) \right)^{k-1}}{(k-1)!} \ e^{-N(F_\nu)}\,,
\end{equation}
with $N(F_\nu)$ representing the number of sources with a flux above $F_\nu$. For details on the derivation and verification hereof see Appendix \ref{app1}. 

\begin{figure}[t!]
\includegraphics[width=\linewidth]{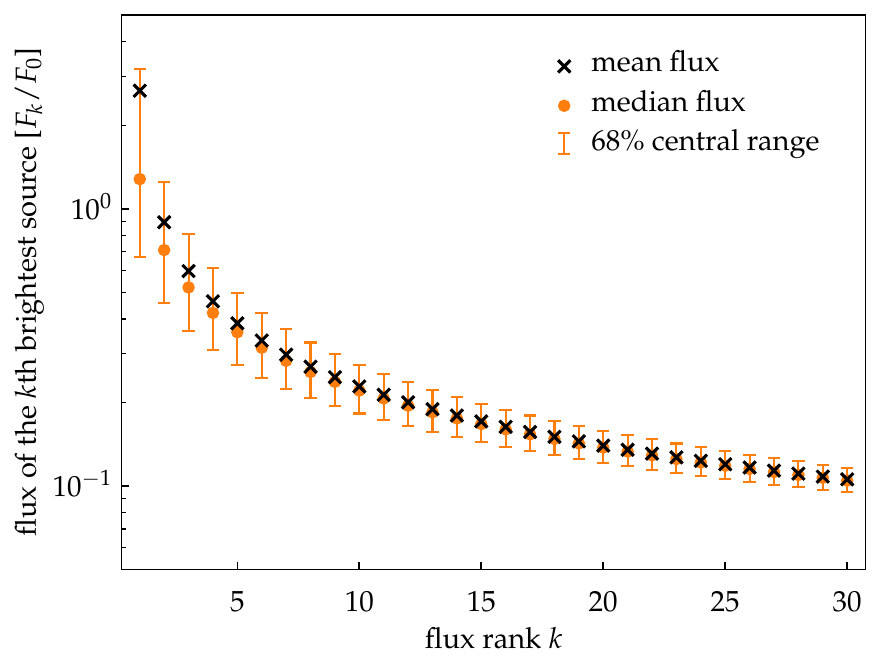}
\caption{Expected flux of the $k$th brightest source following from a distribution in Euclidean space in Eq.~(\ref{eq:dNdF}). The bullet symbols and error bars indicate the median and central $68\%$ range following from the distributions in Eq.~(\ref{eq:p_k}). The crosses show the expectation value from Eq.~(\ref{eq:FkEuclid}).}\label{fig:acc_w}
\end{figure}

In the case of standard-candle populations, we can evaluate the expected flux from the $k$th brightest source using  Eqs.~(\ref{eq:dNdF}) and (\ref{eq:NF}) as:
\begin{equation}\label{eq:FkEuclid}
F_k = \int dF_\nu \, F_\nu \, p_k(F_\nu) = F_0\frac{\Gamma(k-2/3)}{(k-1)!}\,.
\end{equation}
Figure~\ref{fig:acc_w} shows this expected flux for the $30$ brightest sources of a population (crosses) in comparison to the median and 68\% central range of Eq.~(\ref{eq:p_k}). As already mentioned in Section~\ref{sec3}, the brightest source of the population has an expectation value of $F_1 = \Gamma(1/3)F_0 \simeq 2.7\,F_0$, which is close to the upper limit of the 68\% central range. The probability of finding the closest source above a flux level $F_\nu$ is simply $P(F_\nu) = 1-e^{-N(F_\nu)}$, which agrees with our earlier discussion on ensemble variations on the exclusion contours. The median flux of the brightest source is therefore $F_{1,\rm med} \simeq 1.3\,F_0$.

\subsection{Weighted Stacking Searches}

To estimate the relative improvement of the discovery potential for a stacking analysis with respect to that of the analysis of only the brightest source we consider a simple binned maximum-likelihood test, where the bin size is set by the typical angular resolution of neutrino telescopes reaching $0.1^\circ$. The likelihood describing the distribution of events $n_i$ in each of $N_{\rm bin}$ bins is expected to follow from Poisson distributions with mean $\mu_i$:
\begin{equation}\label{eq:LH}
\mathcal{L} = \prod_{i=1}^{N_{\rm bin}} \frac{\mu_i^{n_i}}{n_i!}e^{-\mu_i}\,.
\end{equation}
For the background case (null hypothesis) we simply assume that the expectation values follow a constant background level $\mu_i = \mu_{bg}$. In the case of the signal hypothesis we assume that the first $N_{\rm cat}$ bins are organized such that they follow the (known) location of the brightest sources of a catalogue in ascending order. Defining the stacking weights $w_i \equiv F_i/F_1$ using Eq.~(\ref{eq:FkEuclid}), the expectation values in the signal case can be written as:
\begin{align}\label{eq:mu_i}
	\mu_i = \begin{cases}
		\mu_{\rm sig} w_i + \mu_{\rm bg} & \ \text{for } i\leq N_{\rm cat}\,, \\
		\mu_{\rm bg} & \ \text{for } i > N_{\rm cat}\,,
	\end{cases} 
\end{align}
where $\mu_{\rm sig}$ are the expected events from the brightest source.  

To estimate the discovery potential of $\mu_{\rm sig}$, we define our test statistic (TS) as the maximum log-likelihood-ratio between signal and null hypothesis. Typically, the background level $\mu_{\rm bg}$ for point-source observations is relatively well determined, {\it e.g.}~by off-source scrambling of observed data. The stacking TS is then simply given by:
\begin{equation}\label{eq:TS1}
{\rm TS}_1 \equiv -2 \sum_{i=1}^{N_{\rm cat}}\left(n_i\ln\left(1+\frac{\mu_{\rm sig}w_i}{\mu_{\rm bg}}\right) -\mu_{\rm sig}w_i\right)\,.
\end{equation}
The contribution of the brightest source in the population, $\mu_{\rm sig}\geq0$, is here the only free parameter; the background hypothesis corresponds to $\mu_{\rm sig}=0$. In the limit of large event numbers, the TS distribution under the background hypothesis is expected to follow Wilks' theorem~\cite{Wilks:1938dza}. As the signal strength under the null hypothesis is on the boundary of the physical parameter range, $\mu_{\rm sig}\geq 0$, the tail of the ${\rm TS}_1$ distribution is expected to follow a one-dimensional $\chi^2$ distribution for half of the simulated event distributions under the null hypothesis~\cite{Chernoff:1954eli}; the other half will simply yield ${\rm TS}_1=0$. 

We can now estimate the discovery potential $\mu^{\rm DP}_{\rm sig}$ using {\it Asimov} data sets~\cite{Cowan:2010js}, {\it i.e.}~assuming that the data corresponding to the $5\sigma$ discovery potential can be estimated by the expectation value: $n_i = \mu^{\rm DP}_{\rm sig}{w_i}+ \mu_{\rm bg}$. Simultaneously, we can determine with the same method the discovery potential of a search for {\it only} the closest source, corresponding to $N_{\rm cat}=1$. The relative ratio of the discovery potentials on $\mu_{\rm sig}$ will allow us to estimate the relative increase/decrease of $F^{\rm DP}_1$ in a stacking analysis. See Appendix \ref{app2} for further details on using the Asimov approximation.

\begin{figure*}[t!]
\includegraphics[width=0.49\linewidth]{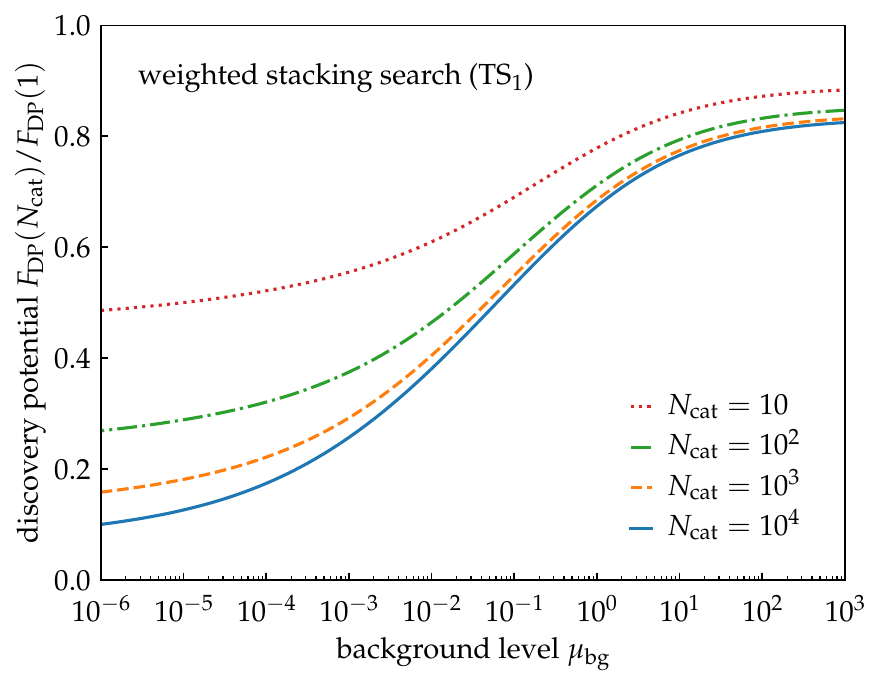}\hfill\includegraphics[width=0.49\linewidth]{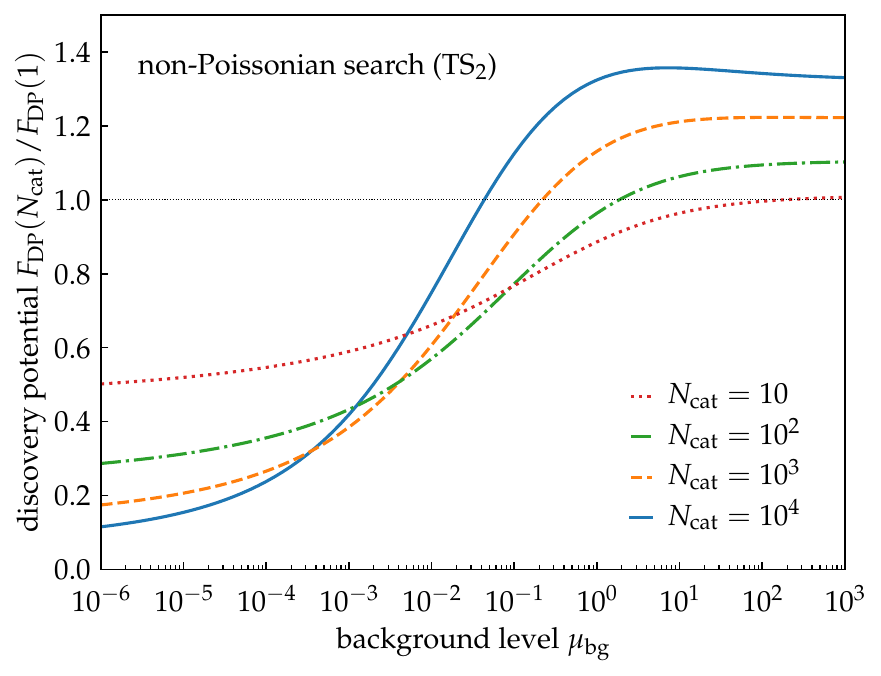}
\caption{Discovery potential of $F_{\rm DP}(N_{\rm cat})$ of population analyses involving the $N_{\rm cat}$ brightest neutrino sources relative to that of only the brightest source for different background levels $\mu_{\rm bg}$. The left plot shows the expected improvements for a weighted stacking search using the test statistic in Eq.~(\ref{eq:TS1}) and the right plot the corresponding improvement for a search of non-Poissonian fluctuations in event-count histograms using Eq.~(\ref{eq:TS2}).}\label{fig:relDP}
\end{figure*}

The left plot of Fig.~\ref{fig:relDP} shows the discovery potential $F_{\rm DP}(N_{\rm cat})$ of the stacking analysis with $N_{\rm cat}$ brightest sources relative to that of just the brightest source. We assume four different numbers of catalogue members and vary the background level $\mu_{\rm bg}$ in Eq.~(\ref{eq:TS1}) in broad range $10^{-6}$ to $10^3$. For high background levels $\mu_{\rm bg} \gg 1$, the relative gain from the stacking search is moderate, reducing the DP to $0.7-0.8$. On the other hand, by assuming very low background levels $\mu_{\rm bg} \ll 1$ the gain can be significant, reaching an improvement by a factor $10$ in the case of a large number of sources.
Similar levels of improvement were recently also discussed in Refs.~\cite{Kowalski:2025iwj,Bartos:2025ooz}.

The main background in the analysis of neutrino point-source emissions comes from atmospheric muons and neutrinos produced by cosmic-ray showers. While atmospheric muons can be efficiently reduced by selecting only upgoing events traversing the Earth, the atmospheric neutrinos produce an irreducible background that reaches the level of IceCube's diffuse astrophysical flux at the pivot energy of $100$~TeV. In case of the IceCube analysis of muon neutrinos based on ten years of data~\cite{IceCube:2019cia}, we can estimate the effective background level $\mu_{\rm bg} \simeq T_{\rm obs}\Delta\Omega A_{\rm eff}\phi_{\rm bg}$ at $100$~TeV assuming a bin size corresponding to the solid angle $\Delta\Omega \simeq \pi(0.3^\circ)^2$, observation time $T_{\rm obs}=10$~years and an effective area of $A_{\rm eff} \simeq 100~{\rm m}^2$ to be at the level of $\mu_{\rm bg} \simeq 10^{-3}-10^{-2}$. For this reference value, we can estimate that stacking searches allow for improvement of the discovery potential of the population by a factor 2--3.

\subsection{Non-Poissonian Fluctuations}

The previous discussion of stacking limits assumed optimal conditions in which the relative weights $w_i$ of the $N_{\rm cat}$ brightest sources can be clearly associated with individual sources in the catalogue. If this association is not possible, {\it e.g.}~due to the unknown relation of the neutrino luminosity to properties of the sources, it is still possible to identify the population of weak neutrino sources by their non-Poissonian fluctuations in event-count histograms~\cite{Malyshev2011,Lee:2015fea,Zechlin:2015wdz,Lisanti:2016jub,IceCube:2019xiu}.

We consider again the optimal situation where the catalogue is complete and contains the $N_{\rm cat}$ brightest sources of the population with relative contribution $F_k$ as in Eq~(\ref{eq:FkEuclid}). Assuming that all sources are separable into $N_{\rm bin} = N_{\rm cat}$ individual bins with background level $\mu_{\rm bg}$, we can define the probability of observing $m_i$ events in a bin $i$ as:
\begin{equation}\label{eq:Pm}
P_{m_i} = \frac{1}{N_{\rm cat}}\sum_{j=1}^{N_{\rm cat}}\frac{\mu_j^{m_i}}{m_i!} e^{-\mu_j}\,.
\end{equation}
The likelihood of observing $n_m$ bins containing $m$ events is then given as:
\begin{equation}
\mathcal{L} = \prod_{i=1}^{N_{\rm cat}}P_{m_i} = \prod_{m=0}^\infty (P_{m})^{n_m}\,.
\end{equation}
Under the background hypothesis we simply have $\mu_j = \mu_{\rm bg}$ and Eq.~(\ref{eq:Pm}) reduces to a Poisson distribution for $m_i$ events. However, under the signal hypothesis with $\mu_j = \mu_{\rm sig}w_j +\mu_{\rm bg}$ the presence of the source population will introduce non-Poissonian fluctuations on top of the background that can be tested in a maximum-likelihood ratio test using the number of bins $n_m$ with event counts $m$.

Analogous to the case of weighted stacking searches, we assume that the background level $\mu_{\rm bg}$ is known for both signal and background hypothesis. The TS of the maximum log-likelihood-ratio becomes here:
\begin{equation}\label{eq:TS2}
{\rm TS}_2 = -2\sum_{m=0}^\infty n_m \ln\Bigg[\sum_{j=1}^{N_{\rm cat}}\left(1+\frac{\mu_{\rm sig}w_j}{\mu_{\rm bg}}\right)^m \frac{e^{-\mu_{\rm sig}w_j}}{N_{\rm cat}}\Bigg] \,,
\end{equation}
where $\mu_{\rm sig}$ is the signal associated with the brightest source of the population. Following the same procedure as before, we can estimate the relative decrease/increase of the discovery potential of the brightest source using Asimov data sets where now the number of bins with event counts $m$ is replaced by the expectation value: $n_m = N_{\rm cat}P_m$. Note that in the limit $N_{\rm cat}=1$, the likelihood in Eq.~(\ref{eq:TS2}) reduces to the one in Eq.~(\ref{eq:TS1}).

The right plot of Fig.~\ref{fig:relDP} shows the discovery potential $F_{\rm DP}(N_{\rm cat})$ for the non-Poissonian search for four different values of the catalogue size $N_{\rm cat}$. For high background levels and a large number of sources the discovery potential is expected to be worse than that of the search from the brightest source. This is to be expected since the weights $w_k$ are here not associated with individual sources and the signal from the brightest source is buried in the large combined background. On the other hand, for low background levels we see improvements to the discovery potential comparable to (but note quite at the level of) those of the weighted stacking search in the left plot of Fig.~\ref{fig:relDP}. Again, assuming an effective background level of $10^{-3}-10^{-2}$ for an energy level of $100~{\rm TeV}$, the discovery potential can be improved by a factor of about 2--3.

\section{\label{sec5} Conclusion}

We discussed in this paper the implications of the non-detection of extragalactic high-energy neutrino point sources (with more than $5\sigma$ significance) for the combined quasi-diffuse emission of their population. This comparison allows us to constrain the contribution of extragalactic populations to the high-energy diffuse flux observed by IceCube with unknown origin.

We reviewed constraints of extragalactic source populations based on their general neutrino luminosity function. Assuming standard-candle populations, these lead to constraints on the neutrino luminosity $L^\star_\nu$ and local source density $\rho_0$, depending on the redshift evolution of their comoving number density. For instance, for neutrino sources with an $E^{-2}$ spectrum evolving with the star-formation rate we derive limits on the (monochromatic) neutrino luminosity at 100~TeV of $L^\star_\nu\gtrsim2\times 10^{44}\,{\rm erg}\,{\rm s}^{-1}$ and density $\rho_0\lesssim 4 \, {\rm Gpc}^{-3}$.

In contrast to earlier studies, we discussed constraints from neutrino point-source studies in terms of their declination-dependent discovery potential and critically assessed the effect of ensemble fluctuations on the exclusion limits. We argue that our method provides more conservative and statistically more robust exclusion limits in the $\rho_0$-$L^\star_\nu$ parameter space compared to studies based on the expected flux from the brightest source and approximate treatment of discovery potentials.

\begin{figure}[t!]
\includegraphics[width=1.\linewidth]{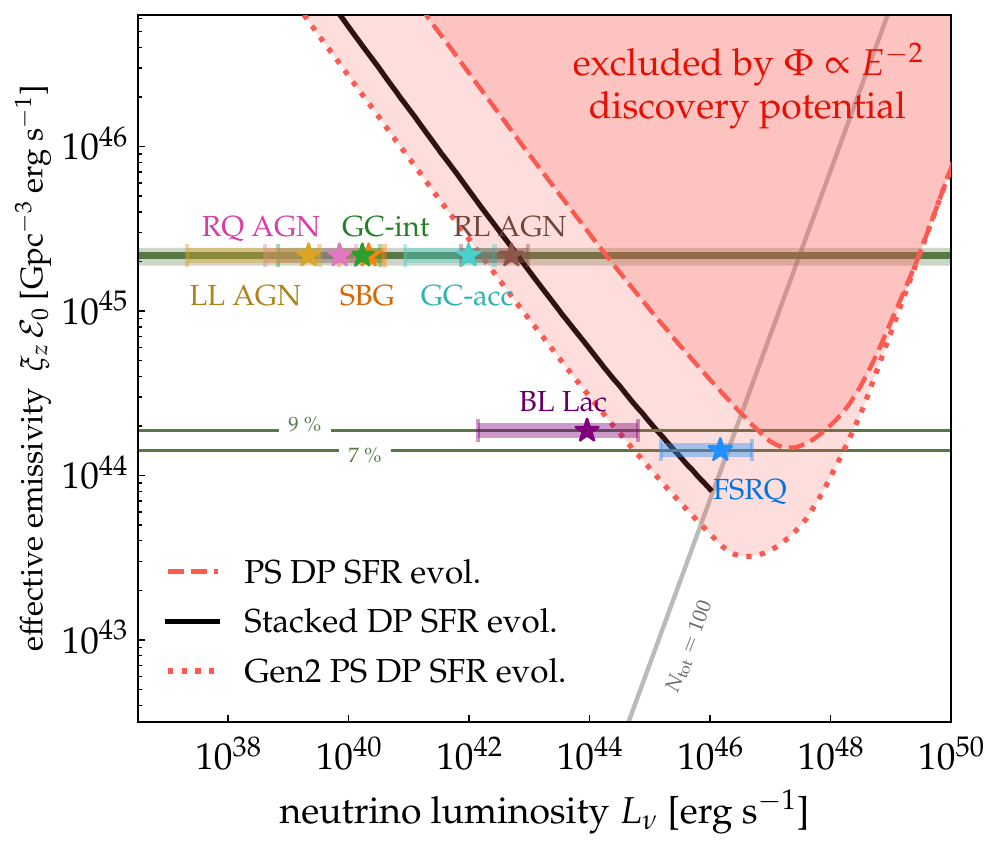}
\caption{Comparison of the exclusion contours based on IceCube's 10 yr point-source discovery potential~\cite{IceCube:2019cia} (dashed red line) to that expected for IceCube-Gen2~\cite{Gen2_TDR} (dotted red line) over the same period. We also indicate the potential reach of catalogue searches (solid black line), assuming a factor 3 improvement compared to IceCube's DP for the brightest source; see Fig.~\ref{fig:rhoL} and Section~\ref{sec4} for details. 
 }\label{fig:stacked}
\end{figure}

We discussed specific candidate populations of steady extragalactic neutrino sources in terms of proxy photon luminosity functions motivated by multi-messenger relations. We critically assessed the applicability of the standard-candle approximation to limit these populations based on their peak neutrino luminosity and corresponding local density. We found that populations where the neutrino luminosity function is not (at least approximately) separable in terms of redshift evolution and luminosity distributions, {\it e.g.}~BL Lacs, require a more detailed treatment. However, we confirmed that high-luminosity candidate sources, such as BL~Lacs and FSRQs, are strongly constrained by point-source limits, consistent with earlier results.

We discussed extensions to previous studies based on the combined analysis of bright neutrino sources of the population, rather than focussing on the brightest source. Using the two examples of a weighted stacking and non-Poissonian fluctuation analyses, we showed that the discovery potential relative to the brightest source of the population can be improved by a factor of 2--3, assuming sufficiently low background levels. This level of improvement would translate into an improvement of the luminosity limit by as much as $3^3 = 27$ as indicated as the black line in Fig.~\ref{fig:stacked}. In addition, the proposed future observatory IceCube-Gen2~\cite{Gen2_TDR} with a point-source discovery potential reduced by a factor 5 compared to IceCube would improve the luminosity limits by factors of $>100$ or even $>1000$ in population analyses.

Finally, our methods outlined in Sections~\ref{sec1} and \ref{sec2} can also be applied to the case of transient candidate neutrino sources such as $\gamma$-ray bursts, tidal disruption events, or flaring AGNs, with minimal modifications; see {\it e.g.}~Refs.~\cite{Ackermann:2019ows,Gen2_TDR, Guepin:2022qpl, IceCube:2022ham, IceCube:2018omy, Yoshida:2022idr}. We expect that the corresponding exclusion limits on the transient source energy and rate will be more conservative than in earlier studies, but leave a discussion for future work.

\begin{acknowledgments}
We thank Kohta Murase for valuable discussions. M.A. and K.M.G.~acknowledge support by Villum Fonden (No.~18994).
\end{acknowledgments}

\appendix

\vspace{-0.3cm}

\section{\label{app1} Flux Distribution of Brightest Sources}

We start from the (arbitrary) distribution $dN/dF$ for the sources of a population in terms of their flux $F$. If $N$ is the total number of sources we can define the probability distribution:
\begin{equation}
p(F) \equiv \frac{1}{N}\frac{dN}{dF}\,,
\end{equation}
of finding a source with flux $F$. The probability $P$ that the flux of a source is greater than a threshold $F$ is defined as:
\begin{equation}
P = \int_{F}^{\infty}dF' p(F')\,.
\end{equation}
The number of sources with flux greater than a threshold $F$ is $N(F) = NP$.

\begin{figure}[t!]
\includegraphics[width=\linewidth]{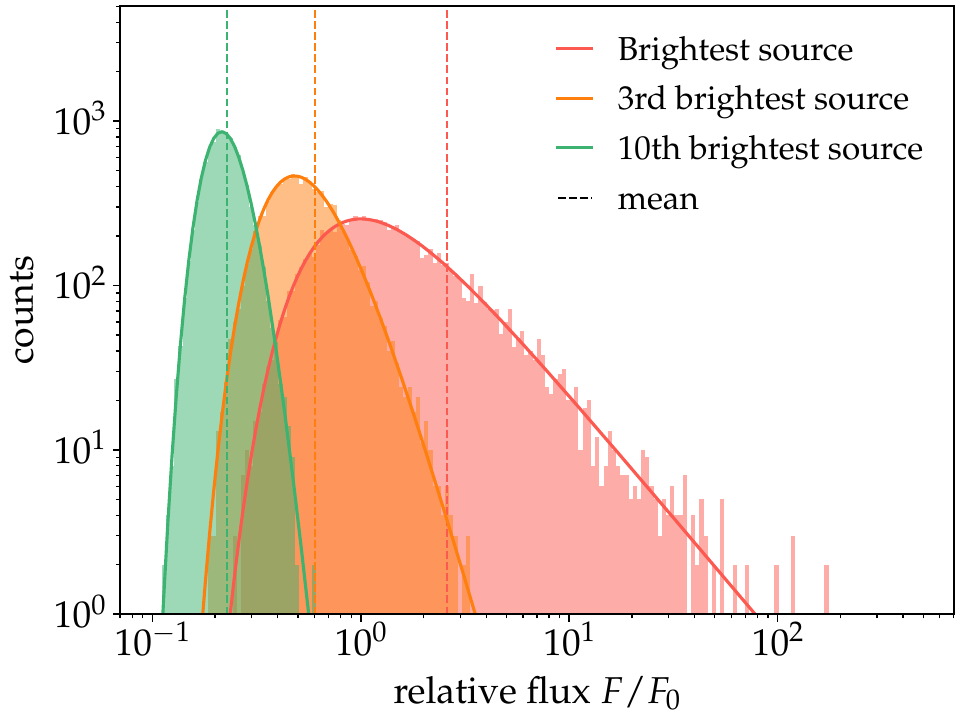}
\caption{\label{fig:p_k} Flux distribution of the brightest, 3rd brightest and 10th brightest source. The solid lines show predictions from Eq.~(\ref{eq:p_k}) and histograms the outcome of $10^4$ simulations for $N_{\mathrm{tot}}=10^6$ sources with flux sampled from Eq.~(\ref{eq:dNdF}). Dashed lines show the median for each distribution.}
\end{figure}

We can then construct the expected flux distribution $p_k(F)$ for $k$th-brightest of the $N$ sources (where $k=1$ defines the brightest source) as follows:
\begin{align}
p_k(F) &= Np(F)\begin{pmatrix}
N-1 \\ k -1
\end{pmatrix}  P^{k-1}  (1-P)^{N-k} \,.
\end{align}
This equation accounts for the cases that any source out of $N$ with individual flux distributions $p(F)$ can play the role of the $k$th-brightest source if any combination of $k-1$ sources out of the remaining $N-1$ sources have a flux larger than $F$. We can rewrite this equation as:
\begin{align}
p_k(F) 
&=   \frac{dN}{dF} \frac{N!}{N^k(N-k)!} \frac{(N(F))^{k-1}}{(k-1)!}  (1-P)^{N-k}\,.
\end{align}
In the large-$N$ limit we get $N!/(N-k)!N^k \to 1$ and $(1-P)^{N-k} \to e^{-N(F)}$, which reproduces the form of Eq.~(\ref{eq:p_k}); see also Ref.~\cite{Ahlers:2014ioa}.

In the large-$N_{\rm tot}$ limit, the distributions $p_k(F)$ are appropriately normalized to $\int dF p_k(F) = 1$. In addition, we recover the full flux distribution via:
\begin{equation}
\sum_k p_k(F) = \frac{dN}{dF}\,.
\end{equation}
Figure~\ref{fig:p_k} illustrates the flux distributions $p_k(F)$ for the cases $k=1$, $k=3$, and $k=10$ together with the result of Monte Carlo simulations. We repeatedly simulated the flux of $N=10^6$ neutrino sources following the distribution of Eq.~(\ref{eq:dNdF}) to determine the flux distribution of the $k$-th brightest source. The resulting histograms show excellent agreement with the prediction from Eq.~(\ref{eq:p_k}).

\vspace{-0.3cm}

\section{\label{app2} Asimov Approximation}

In Section~\ref{sec4} we used Asimov data sets to estimate the DP of weighted stacking and non-Poissonian fluctuation analyses relative to the DP of the brightest source, shown in Fig.~\ref{fig:relDP}. In this section, we validate that this method is a good approximation even in the case of low background and signal levels by determining the $3\sigma$ DP from TS distributions inferred from simulated data.

\begin{figure*}[t!]
\includegraphics[width=0.85\linewidth]{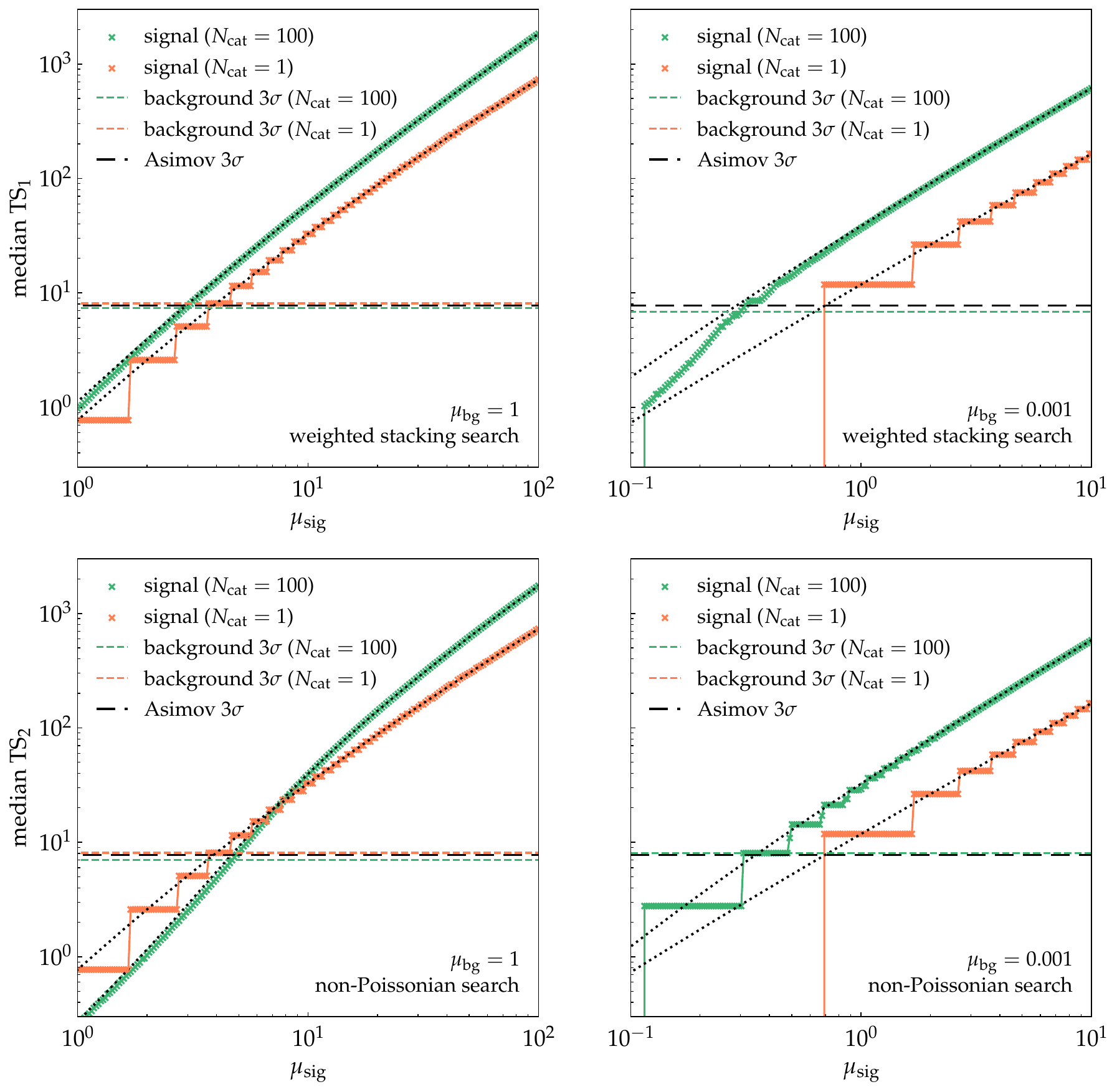}
\caption{\label{fig:med} Median test statistic for weighted source stacking (${\rm TS}_1$; top panels) and non-Poissonian fluctuations (${\rm TS}_2$; bottom panels) assuming a catalogue of the $N_{\rm cat}=100$ brightest sources (green) compared to just the brightest source, $N_{\rm cat} = 1$ (orange). The results are shown for the two background levels $\mu_{\rm bg}=1$ (left panels) and $\mu_{\rm bg}=10^{-3}$ (right panels). The crosses show the results of simulations with increasing signal level $\mu_{\rm sig}$ that follow the trend of the superimposed dotted lines derived from the Asimov approximation. The narrow-dashed horizontal lines show the $3\sigma$ discovery potential inferred from background simulations, whereas the wide-dashed horizontal lines show that used for the Asimov approximation.}\label{fig:Asimov}
\end{figure*}

For each test statistic, we determine the $3\sigma$ TS threshold by simulating background data sets and by maximizing the TSs in Eqs.~(\ref{eq:TS1}) or (\ref{eq:TS2}) for either $N_{\rm cat}=1$ or $N_{\rm cat}=100$. The resulting thresholds are indicated as orange ($N_{\rm cat}=1$) and green ($N_{\rm cat}=100$) horizontal dashed lines in Fig.~\ref{fig:Asimov} for ${\rm TS}_1$ (top panels) and ${\rm TS}_2$ (bottom panels) for a per-bin background level $\mu_{\rm bg}=1$ (left panels) and $\mu_{\rm bg}=10^{-3}$ (right panels). In the case $N_{\rm cat}=1$ and $\mu_{\rm bg}=10^{-3}$ the simulated $3\sigma$ threshold is at $0$ and is not shown in the plots. 

On the other hand, the Asimov approximation assumes that the null hypothesis follows Wilks' theorem~\cite{Wilks:1938dza}. Accounting for the fact that the null hypothesis $\mu_{\rm sig}=0$ falls on the boundary of the signal values, $\mu_{\rm sig}\geq0$, we expect here that half of the background data follows the one-dimensional $\chi^2$ distribution~\cite{Chernoff:1954eli}; the other half will evaluate to $\mathrm{TS}=0$. The corresponding $3\sigma$ threshold is indicated as the black dashed horizontal lines in Fig.~\ref{fig:Asimov}.

The required $3\sigma$ DP of the signal strength $\mu_{\rm sig}$ (normalized to the brightest source) is then determined from the median of TS distributions from $10^5$ signal simulations with logarithmically increasing values of $\mu_{\rm sig}$. The results are shown in Fig.~\ref{fig:Asimov} as the coloured crosses for the cases $N_{\rm cat}=1$ (orange) and $N_{\rm cat}=100$ (green). The required $3\sigma$ level of $\mu_{\rm sig}$ is determined by the crossing of the horizontal dashed lines.

In the case of the Asimov approximation, the median of the TS distribution is replaced by the TS value of Asimov data that follows from the expectation values used in the simulation; for details, see Section~\ref{sec4}. The corresponding TS values are superimposed on top of the median of the simulated TS distributions in Fig.~\ref{fig:Asimov} as dotted lines. Besides fluctuations that are related to the discreteness of simulated data, we find good agreement between the two methods. The average relative improvement of the population analyses is well reproduced for the eight different combinations of $N_{\rm cat}=1$ or $100$, ${\rm TS}_1$ or ${\rm TS}_2$ and $\mu_{\rm bg}=10^{-3}$ and $1$. Note that in the case $N_{\rm cat}=1$, both TSs become equivalent, which is reproduced by the corresponding orange histograms in the upper and lower panels.

\bibliographystyle{utphys_mod}
\bibliography{references}

\end{document}